\documentclass[final,3p,times]{elsarticle}

\usepackage{amssymb}
\usepackage{amsmath}
\newcommand{\eos}{EoS}

\journal{Journal of Subatomic Particles and Cosmology}

\begin{document}

\begin{frontmatter}



\title{Nuclear matter equation of state and astrophysics}

\author[mrp]{Mateus Reinke Pelicer}
\affiliation[mrp]{organization={University of Illinois Urbana--Champaign},
             addressline={601 East John Street},
             city={Champaign},
             postcode={61801},
             state={Illinois},
             country={USA}}

\begin{abstract}
Neutron-star masses, radii, and inspiral tidal deformabilities now provide quantitative constraints on the cold equation of state (\eos), favoring relatively soft matter around one to two times nuclear saturation density and substantial stiffening at larger density. These bulk constraints, however, do not uniquely determine the microscopic composition of the stellar core. Hyperons, deconfined quarks, quarkyonic matter, and strong first-order phase transitions remain viable possibilities. This article summarizes the present multimessenger status and emphasizes the next challenge---a unified description of strongly interacting matter across catalyzed neutron stars, binary mergers, and heavy-ion collisions. Recent results presented at SQM2026, including new constraints on hyperon interactions and advances in multidimensional equation-of-state modeling, highlight the complementary experimental and theoretical inputs required for this program. The MUSES Calculation Engine provides modular software infrastructure for connecting these inputs to astrophysical and heavy-ion applications.
\end{abstract}

\begin{keyword}
neutron-star \sep dense-matter \sep hyperons



\end{keyword}

\end{frontmatter}


\section{Equation of state constraints from compact stars}

A neutron star compresses roughly $1$--$2.3\,M_\odot$ into a radius of order $10$--$14$ km, reaching several times the nuclear saturation density $n_0$ at its core. Cold, catalyzed compact stars are well approximated as cold, neutrino-transparent, charge neutral, and in weak equilibrium. Once a barotropic equation of state (\eos) $P(\varepsilon)$ is specified, their macroscopic structure follows from the Tolman--Oppenheimer--Volkoff equations.
Consequently, measurements of masses, radii, and tidal responses constrain the \eos, although the inverse map from a finite set of observations to microscopic matter is non-unique~\cite{Watts:2014tja,Lattimer:2016hyd}.

The first robust astrophysical requirement is an \eos\ sufficiently stiff to support pulsars with $2\,M_\odot$ or above~\cite{Antoniadis:2013pzd,Fonseca:2021wxt}. NICER pulse-profile modeling has complemented these mass measurements with simultaneous mass-radius constraints, generally favoring radii of order $12$--$13$~km\cite{Riley:2021pdl,Miller:2019cac,Kini:2026rjx}. GW170817 independently provided the first gravitational-wave constraint on the binary tidal deformability, disfavoring very large stellar radii and limiting the \eos\ pressure at intermediate densities~\cite{LIGOScientific:2017vwq}.

Theoretical information provides complementary constraints at the boundaries of the density range relevant to neutron stars.
Chiral effective field theory constrains the \eos\ around saturation density, with systematic uncertainties~\cite{Drischler:2021bup,Drischler:2021kxf}. 
At asymptotically high chemical potential, perturbative QCD supplies an additional anchor, and causality and thermodynamic stability propagate that information toward neutron-star densities~\cite{Komoltsev:2021jzg,Gorda:2022jvk}. 
Flexible nonparametric or weakly parametric inference frameworks are useful between these regimes because they reduce restrictive correlations in density imposed by specific functional forms~\cite{Landry:2018prl,Essick:2019ldf,Capano:2019eae,Annala:2021gom,Legred:2021hdx,Legred:2022pyp}.

Although the quantitative conclusions depend on the chosen dataset and the prior, a recurring picture is relatively soft matter around $n_0$--$2n_0$, followed by substantial stiffening at larger density. This behavior can be expressed through the equilibrium sound speed
\begin{equation}\label{eq:cs2}
 c_s^2=\left.\frac{dP}{d\varepsilon}\right|_{\rm eq},
\end{equation}
which several analyses find must exceed the conformal value $1/3$ somewhere inside compact stars\cite{Bedaque:2014sqa,Tews:2018kmu,Mroczek:2023zxo}. 
Nonmonotonic structures in $c_s^2$ may reflect changes in composition, interactions, or phase structure, but their microscopic origin cannot generally be inferred from the bulk \eos\ alone. Distinct microscopic models can generate similar $P(\varepsilon)$ relations and nearly indistinguishable mass--radius curves. 

\section{Composition as the central uncertainty}

Hyperons are energetically favored when the baryon density becomes sufficiently large, but their appearance softens the \eos\ in many traditional models and reduces the maximum neutron-star mass supported by it. Reconciling hyperons with two-solar-mass stars constitutes the hyperon puzzle~\cite{Chatterjee:2015pua,Sedrakian:2022ata}. Its resolution depends on poorly known hyperon--nucleon and hyperon--hyperon interactions in dense matter and may require additional repulsion, potentially including three-body forces. 

This is an area where heavy-ion measurements directly inform astrophysics. At SQM2026, the STAR collaboration, represented by Jing Gu, presented measurements of the three-particle $pp\Lambda$ correlation in Au+Au collisions at $\sqrt{s_{NN}}=3$ GeV, close to the hyperon-production threshold. Henrik Fribert presented new ALICE measurements of $p\Sigma^-$ and $p\Sigma^+$ correlations, while Raffaele Del Grande discussed the ALICE program probing $\Lambda pp$ and $\Xi^-pp$ three-body dynamics. More generally, two- and three-particle correlation measurements provide access to interactions involving unstable hyperons, for which direct scattering information is scarce~\cite{Fabbietti:2020bfg,ALICE:2020mfd,Vidana:2024ngv}. Although these measurements do not determine the neutron-star \eos\ by themselves, they help constrain hyperon--nucleon and hyperonic three-body interactions that constitute important microscopic inputs to \eos\ models.

The presence of quark degrees of freedom is equally difficult to establish observationally. Hybrid and quarkyonic constructions can satisfy present mass, radius, and tidal constraints, but so can purely nucleonic descriptions\cite{McLerran:2018hbz,Alford:2019oge,Fujimoto:2024doc}. Thus, neither the presence nor the absence of quark matter is presently inferred in a model-independent way. Strong first-order transitions can produce disconnected stellar branches or sharp features in the sound speed, but smooth crossovers and weak transitions remain observationally degenerate with them.

Binary neutron-star postmergers offer a qualitatively new probe. The dominant postmerger frequency and collapse time depend on the finite-temperature  \eos\ out of weak equilibrium, and sufficiently strong first-order transitions can leave characteristic signatures~\cite{Most:2018eaw,Ripley:2023lsq,Mroczek:2024sfp}. Their interpretation remains challenging because thermal effects, composition-dependent weak rates, and numerical relativity systematics can mimic or obscure phase-transition signals. Nevertheless, future observations may provide substantially better sensitivity to the microscopic composition rather than only to the cold bulk pressure.

\section{Multidimensional \eos\ and nonequilibrium dynamics}

For cold catalyzed matter, weak equilibrium sets the strangeness chemical potential to zero and relates the electric charge chemical potential to the leptonic sector. Heavy-ion collisions, by contrast, conserve net baryon number, strangeness, and electric charge on strong-interaction time scales. During the merging of compact objects, neutron stars can be driven temporarily away from weak equilibrium. Therefore, a common description must start from the multidimensional thermodynamic potential
\begin{equation}
 P=P(T,\mu_B,\mu_S,\mu_Q),\qquad
 n_X=\frac{\partial P}{\partial\mu_X},\quad
 s=\frac{\partial P}{\partial T},
\label{eq:4d}
\end{equation}
with $X\in\{B,S,Q\}$, subject to appropriate constraints to each system. 

The multidimensional viewpoint was a major theme of SQM2026. Grefa \emph{et al.} extracted chemical-potential differentials from Ru+Ru and Zr+Zr isobar data and compared them with lattice-QCD and chiral mean-field predictions~\cite{Grefa:2026meq}. Yang presented a generalized symmetry-energy expansion for strange dense matter, providing a controlled bridge between heavy-ion and neutron-star conditions~\cite{Yang:2025wop}. Vovchenko showed a density-dependent interacting hadron-resonance gas constrained by lattice susceptibilities, nuclear matter, and neutron-star observations~\cite{Fujimoto:2021dvn,Moss:2024uam}. Kahangirwe discussed an extended chiral mean-field description with medium-modified strange and non-strange mesons~\cite{Kumar:2025rxj}.

The phase structure also becomes richer away from the usual $(T,\mu_B)$ plane. Jahan presented an extension of the $T'$-expanded lattice-QCD \eos\ to a parametrized critical surface at finite $(\mu_B,\mu_Q,\mu_S)$~\cite{Jahan:2025vwa,Abuali:2025tbd}. Noronha-Hostler discussed chiral mean-field parameterizations containing, besides liquid--gas and deconfinement transitions, a strangeness-dominated phase and a third critical point~\cite{Cruz-Camacho:2024odu}. Such structures are model dependent, but they demonstrate why fixing $\mu_S$ or $\mu_Q$ implicitly can bias predictions.

In a merger, fluid elements oscillate and compress on millisecond time scales, while strong, electromagnetic, and weak processes equilibrate different charges on widely separated time scales. If weak reactions are fast, the composition follows its equilibrium trajectory and the restoring force is governed by Eq.~(\ref{eq:cs2}). If reactions are slow, particle fractions are effectively frozen and the corresponding sound speed is generally larger than the equilibrium one. The difference between these limits drives chemical relaxation and bulk viscosity.

When several leptonic, nonleptonic, hyperonic, or quark reaction channels coexist, the response is characterized by multiple equilibration times that control the damping of fluid oscillations: density perturbations generate several independent chemical imbalances, whose relaxation depends jointly on microscopic reaction rates and on the susceptibilities relating density and chemical-potential variations.
Cold stars satisfy $\mu_S=0$ after weak equilibration, but a merger can generate transient $\mu_S\neq0$ and $\mu_Q$ fluctuations before the relevant reactions restore equilibrium. Recent work has begun to evolve such chemical degrees of freedom explicitly and to quantify their impact on merger dynamics and gravitational waves~\cite{Most:2022yhe,Alford:2024tyj}. A four-dimensional \eos\ table must consequently provide accurate first and second derivatives and a proper treatment of metastable and unstable branches. Connecting microscopic models to heavy-ion and numerical relativity simulations requires the numerical infrastructure needed to match, interpolate, invert, and differentiate an \eos\ without violating thermodynamic consistency.

\section{MUSES as community infrastructure}

The Modular Unified Solver of the Equation of State (MUSES) was developed to address these requirements within a modular and reproducible computational framework. Its cyberinfrastructure combines independent, open-source physics modules; common standards for their inputs, outputs, and documentation; and a Calculation Engine that orchestrates them as user-defined workflows~\cite{ReinkePelicer:2025vuh}. Because individual modules can be replaced without reconstructing the complete pipeline, assumptions that are often hidden within \eos\ tables become explicit and testable.

The first public MUSES workflows focused on cold, charge-neutral matter in $\beta$ equilibrium. They connected a crust density-functional description, chiral effective field theory around nuclear saturation, and a chiral mean-field model at high density using a smooth matching technique, and then computed the associated stellar structure using the QLIMR module~\cite{ReinkePelicer:2025vuh}. Varying the thermodynamic variable and density range used for smooth matching changed the radius of a $1.4\,M_\odot$ star by up to approximately $9\%$ and the maximum mass by up to approximately $4\%$. The different matching prescriptions preserve some thermodynamic features while introducing artificial structure into others, particularly into derivatives such as the sound speed. These uncertainties must be regarded as part of the physical modeling. The same modular structure also enables systematic model calibration: at SQM2026, Cruz Camacho presented a Fisher-information analysis identifying the combinations of chiral mean-field parameters to which neutron-star observables are most sensitive~\cite{Cruz-Camacho:2026pdg}.

The Calliope release extends this architecture from cold neutron-star matter to the finite-temperature, multidimensional thermodynamics required for heavy-ion collisions~\cite{Jahan:2026hvs}. It includes hadronic, lattice-QCD-based, and phenomenological \eos\ modules defined over two- to four-dimensional regions of $(T,\mu_B,\mu_S,\mu_Q)$, including models with a tunable critical point. Besides providing standalone tables from specific frameworks, the MUSES framework provides modules to merge different \eos\ and to invert tables from chemical potentials and temperature into variables suitable for hydrodynamic and numerical relativity simulations. At SQM2026, Garella presented a statistical-mixture extension of the Synthesis module in which the merged thermodynamics derives from a single grand potential~\cite{Yang:2026brr}. 

When fluid cells leave the valid range of an expansion-based \eos\ and must be evaluated with a backup prescription, exploratory simulations show that the resulting contamination can be substantial for total multiplicity and, to a lesser extent, elliptic flow, particularly at lower collision energies~\cite{Jahan:2026hvs}. This shows that limited phase-diagram coverage can propagate directly into the observables used to constrain the \eos, particularly in regimes of high baryon density and finite strangeness.

\section{Outlook}

Multimessenger observations have substantially narrowed the range of viable cold equations of state. Massive pulsars require sufficient stiffness at high density, while radius and tidal-deformability measurements favor moderate pressures closer to nuclear saturation. The microscopic interpretation of these constraints remains far less certain.  Resolving this ambiguity requires combining complementary information from astrophysical observations, laboratory measurements of strange interactions, lattice QCD, effective theories, and dynamical simulations. It also requires moving beyond a single cold barotropic relation. Heavy-ion collisions and neutron-star mergers probe matter at finite temperature, with independently evolving baryon number, strangeness, and electric charge. The appropriate framework is therefore the multidimensional thermodynamic potential $P(T,\mu_B,\mu_S,\mu_Q)$, together with its derivatives, phase structure, and process-dependent response.

The developments presented at SQM2026 illustrate the need for common computational infrastructure. Advances in multidimensional modeling, thermodynamic matching, inversion, and uncertainty analysis are making increasingly realistic equations of state available for simulation. MUSES provides a modular framework for using these ingredients consistently in neutron-star and heavy-ion calculations. It allows competing microscopic assumptions to be tested across systems and their modeling uncertainties to be propagated into observable predictions.

\section{Acknowledgments}
M.R.P. is supported by the NSF under the MUSES collaboration OAC2103680.

\bibliographystyle{elsarticle-num}
\bibliography{references}



\end{document}